\documentclass[amsmath,aps]{revtex4}
\usepackage{graphicx,color}
\usepackage{CJK}
\usepackage{bm}
\usepackage[hypertex]{hyperref}

\newcommand{\be}{\begin{equation}}
\newcommand{\ee}{\end{equation}}
\newcommand{\bea}{\begin{eqnarray}}
\newcommand{\eea}{\end{eqnarray}}
\newcommand{\bsube}{\begin{subequations}}
\newcommand{\esube}{\end{subequations}}

\newcommand{\beq}{\begin{equation}}
\newcommand{\eeq}{\end{equation}}
\newcommand{\beqn}{\begin{eqnarray}}
\newcommand{\eeqn}{\end{eqnarray}}
\newcommand{\bsub}{\begin{subequations}}
\newcommand{\esub}{\end{subequations}}

\begin{document}

\title{Band structures of strained Kagome lattices}

\author{Luting Xu}
\affiliation{Center for Joint Quantum Studies and Department of Physics,
School of Science, Tianjin University, Tianjin 300354, China}

\author{Fan Yang}
\email{fanyangphys@tju.edu.cn}
\affiliation{Center for Joint Quantum Studies and Department of Physics,
School of Science, Tianjin University, Tianjin 300354, China}

\date{\today}

\begin{abstract}
Materials with kagome lattice have attracted significant research attention due to their nontrivial features in energy bands. In this work, we theoretically investigate the evolution of electronic band structures of kagome lattice in response to uniaxial strain using both a tight-binding model and an antidot model based on a periodic muffin-tin potential. It is found that the Dirac points move with applied strain. Furthermore, the flat band of unstrained kagome lattice is found to develop into a highly anisotropic shape under a stretching strain along $y$ direction, forming a partially flat band with a region dispersionless along $k_y$ direction while dispersive along $k_x$ direction. Our results shed light on the possibility of engineering the electronic band structures of kagome materials by mechanical strain.
\end{abstract}


\maketitle

\section{Introduction}

The kagome lattice is a two-dimensional (2D) hexagonal Bravais lattice consisting of corner-sharing triangles. It has been discovered as a fertile land for realizing various exotic physics since introduced by Itiro Sy\^{o}zi in 1951\cite{syozi1951}. Due to the frustrated lattice geometry and  nontrivial band features,  kagome lattice is predicted to be a promising platform for investigating novel physics like frustration-driven magnetism\cite{Yin2018, Yin2019}, quantum-spin-liquid states\cite{Lee2007, Balents2010, Yan2011, Jiang2012, Han2012, Norman2016, Liao2017}, topological phases\cite{Guo2009, Chisnell2015, Yin2022} and electron correlations\cite{Zaanen2015}.

Unfortunately, materials with kagome lattice are rare in nature. As an alternative approach to explore the physics related to kagome lattice, various artificial kagome systems, such as optical kagome lattices\cite{Guzman2012, Leung2020, Hassan2019} and monatomic kagome layer grown on metal surfaces\cite{Lin2022}, have been experimentally developed. The advantage of artificial kagome systems is that it allows the individual tuning of all system parameters in a wide range, and thus enables searching for the predicted non-trivial features of kagome lattices in a much larger parameter space. For example, in natural kagome materials, parameters like hopping energy, defect density and mechanical strain are difficult to control experimentally, but they can all be conveniently tuned in artificial kagome systems by simply varying the design and geometry of the artificail lattice.

Mechanical strain has proven to be an important tool for engineering electronic band structures\cite{Roldan2015, Dai2019, Peng2020}. Over the past decades, extensive researches have been carried out on the strain effect in graphene\cite{Pereira2009, Feilhauer2015, Shioya2015, Cao2020}, carbon nanotube\cite{Paulson1999, Minot2003, Dmitrovic2015}, MoS$_2$\cite{Conley2013, He2013, Zhu2013}, WSe$_2$\cite{Desai2014, Parto2021} and many other materials. In contrast, studies of strained kagome systems have just started\cite{Liu2020, Wamg2022, Nayga2022, Liu2019, Wang2022}; a systematic investigation of the band structures of strained kagome lattice is still lacking.

In this work, we theoretically study the electronic band structures of strained kagome lattice using both a tight-binding model and an antidot model based on a periodic muffin-tin potential. The main findings include: (i) The Dirac points of kagome lattice are shifted away from the corners of Brillouin zoom under applied uniaxial strain. In contrast to the situation in strained graphene\cite{Feilhauer2015}, the Dirac cones of strained kagome lattices never merge within the framework of nearest-neighbor tight-binding model. However, in a more realistic model based on a periodic muffin-tin potential, the Dirac cones do merge with increasing compressive strain, causing band-gap openings at the Dirac points. (ii) When a stretching strain is applied in $y$ direction, the flat band of the unstrained kagome lattice becomes highly anisotropic, forming a partially flat band with an area that is dispersionless along $k_y$ direction whereas dispersive along $k_x$ direction.

The paper is organized as follows. In Section 2, we calculate the energy bands of strained kagome lattices using a nearest-neighbor tight-binding model. In Section 3, we first introduce a antidot model for artificial kagome lattice, and then numerically investigate its band structures under uniaxial strain. The results and conclusions are summarized in Section 4.

\section{Tight-binding approach to strained kagome lattice}

 In this section, we investigate the band structures of strained kagome lattice using the tight-binding approximation. Since the band structures of kagome lattice are primarily characterized by the coexistence of Dirac cones and a flat band, here we mainly focus on the shift of the Dirac cones and the reshaping of the flat band in response to applied uniaxial strain.

 \subsection{Tight-binding model for kagome lattice}

 Fig. 1(a) is an illustration of the nearest-neighbor tight-binding model for strained kagome lattice. The kagome lattice is a triangular Bravais lattice with a unit cell composed of three inequivalent sites, which are labeled as $A$, $B$ and $C$, respectively. The three sites in the unit cell form a regular triangle, as indicated by the shaded region in Fig. 1(a). The repeated Brillouin zone of kagome lattice is shown in Fig. 1(b), where the first Brillouin zone is a regular hexagon with two independent corners $K$ and $K^{\prime}$ .

 The tight-binding Hamiltonian of kagome lattice is given as\cite{Guo2009, Liu2020}
\begin{equation}
H=-\sum_{\bm{r}}[t_1(a^\dagger_{\bm{r}}b_{\bm{r}}+a^\dagger_{\bm{r}}b_{{\bm{r}}+\bm{\delta}_1})+t_2(a^\dagger_{\bm{r}}c_{\bm{r}}+c^\dagger_\textbf{r}a_{{\bm{r}}+\bm{\delta}_2})
+t_3(b^\dagger_{\bm{r}}c_{\bm{r}}+c^\dagger_{\bm{r}}b_{{\bm{r}}+\bm{\delta}_3})]+\rm{h.c.}\label{Hum1},
\end{equation}
with the summation of $\bm{r}$ running over all unit cells. Here $a^\dagger_{\bm{r}}$, $b^\dagger_{\bm{r}}$ and $c^\dagger_{\bm{r}}$ ($a_{\bm{r}}$, $b_{\bm{r}}$, and $c_{\bm{r}}$) are respectively the creation (annihilation) operators of electrons at sites $A$, $B$ and $C$ in the unit cell located at $\bm{r}$, and $t_i (i=1,2,3)$ is the hopping parameter along the direction of lattice vectors $\bm{\delta}_i$, which are given by
\begin{equation}
\bm{\delta}_1=a \bm{e}_x+\sqrt{3}a\bm{e}_y, \, \bm{\delta}_2=a \bm{e}_x-\sqrt{3}a\bm{e}_y,\, \textrm{and}~\bm{\delta}_3=2a \bm{e}_x\label{vec1},
\end{equation}
where $\bm{e}_x$ and $\bm{e}_y$ are basis vectors of the Cartesian coordinate system, and $a$ is the distance between neighboring sites.

\begin{center}
  \vspace{+0.2cm}
  \includegraphics[width=0.45 \linewidth]{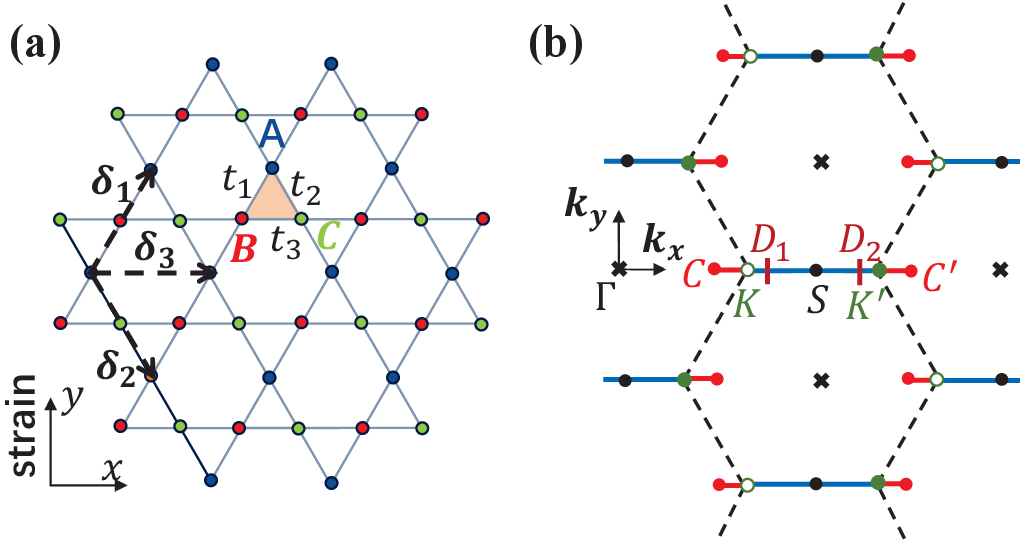}\\[15pt]  
  \parbox[c]{15.0cm}{\footnotesize{\bf Fig.~1.}
   (color online) (a) Tight-binding model of kagome lattice with only nearest-neighbour hopping terms. The three inequivalent sites in the unit cell are labeled as $A$, $B$ and $C$, respectively; $t_i(i=1,2,3)$ denotes the hopping parameter between the nearest neighboring atoms along direction $\boldsymbol{\delta}_i$. A uniaxial strain in $y$ direction is modeled by a group of anisotropic hopping parameters $t_1=t_2=t^\prime$ and $t_3=t$.
   (b) The repeated Brillouin zone of kagome lattice.
   The center point $\Gamma$ and two independent corners $K$ and $K^\prime$ are denoted by the black crosses, hollow green circles and solid green circles, respectively. When the lattice is strained in $y$ direction, the Dirac points $D_1$ and $D_2$ are shifted away from $K$ and $K^\prime$ and located somewhere between $C$ and $C^\prime$.}\label{Fig1model}\\[15pt]
\end{center}

By applying the Wannier transformation
\begin{equation}
\left(
 \begin{array}{c}
 a_{\bm{r}}\\
 b_{\bm{r}}\\
 c_{\bm{r}}
 \end{array}
 \right)
 =\sum_{\bm{k}\in BZ}\exp(i\bm{k}\cdot \bm{r})
 \left(
 \begin{array}{c}
 a_{\bm{k}}\\
 b_{\bm{k}}\\
 c_{\bm{k}}
 \end{array}
 \right)
\end{equation}
to Eq.(\ref{Hum1}), the Hamiltonian can be rewritten in momentum space as $H=\sum_{\bm{k}}\Psi^\dagger_{\bm{k}}\mathcal{H}_{\bm{k}}\Psi_{\bm{k}}$, where $\Psi_{\bm{k}}=(a_{\bm{k}},b_{\bm{k}},c_{\bm{k}})^T$ and the Hamiltonian matrix $\mathcal{H}_{\bm{k}}$ reads
\begin{equation}
\mathcal{H}_{\bm{k}}= \left(
\begin{array}{ccccc}
0 & -t_1(1+e^{ik_1})  & -t_2(1+e^{-ik_2}) \\
 -t_1(1+e^{-ik_1}) & 0 & -t_3(1+e^{-ik_3}) \\
 -t_2(1+e^{ik_2})  & -t_3(1+e^{ik_3})  &0
\end{array}
\right).
\end{equation}
Here $k_i\equiv\bm{k}\cdot \bm{\delta}_i~(i=1,2,3)$, which satisfies $k_1+k_2=k_3$.

In an unstrained kagome lattice, the 6-fold rotational symmetry ensures $t_1=t_2=t_3$. However, as shown in Fig. 1(a), an uniaxial strain applied in $y$ direction breaks the rotational symmetry and consequently leads to $t_1=t_2\neq t_3$. Assuming that $t_1=t_2=t^\prime$ and $t_3=t$, the anisotropy ratio $\alpha$ can be defined as $\alpha\equiv t^\prime/t$. The parameter $\alpha$ characterizes both the type and strength of the uniaxial strain, as explained as follows: (i) A decrease in inter-site distance always causes an increment of hopping parameter and \textit{vise versa}, and hence $\alpha>1$ ($\alpha<1$) corresponds to the situation where a compressive strain (stretching strain) is applied along $y$ direction. (ii) Apparently, the further $\alpha$ deviates from 1, the stronger the applied strain is.

Meanwhile, as a reasonable approximation for weak strain\cite{Feilhauer2015}, we neglect the site shift in the strained lattice and assume that the lattice will keep its shape when $\alpha \neq 1$. Based on this assumption, a uniaxial strain applied in $y$ direction can be modeled using a single parameter $\alpha$, which greatly reduces the mathematical complexity. The eigenvalue equation of $\mathcal{H}_{\bm{k}}$ is consequently reduced to
\begin{equation}\label{EnerTB}
E^3_{\bm{k}}+4\alpha^2t^3 (1+\sum_{i}\cos k_i)-2E_{\bm{k}}t^2(\cos k_1+1)-2E_{\bm{k}}t^2\alpha^2( \cos k_2+ \cos k_3+2)=0.
\end{equation}

 \subsection{Results and discussions}

 \subsubsection{General features of energy bands}
The energy bands of strained kagome lattice can be obtained by solving Eq.(\ref{EnerTB}). Figs. 2(a)-(c) shows the band structures calculated with different values of $\alpha$.

In the unstrained case where $\alpha=1.0$, the solutions of Eq.(\ref{EnerTB}) gives two dispersive bands $E^{(1,2)}_{\bm{k}}=t[-1\pm \sqrt{2\sum_{i}\cos k_i+3}]$, and a 2D flat band $E^{(3)}_{\bm{k}}=2t$. In resemblance to the energy bands of graphene, the two lower dispersive bands $E^{(1)}_{\bm{k}}$ and $E^{(2)}_{\bm{k}}$ intersect at the corners of the hexagonal Brillouin zone, forming two linearly dispersive Dirac cones at $K$ and $K^{\prime}$, as shown in Fig. 2(a). The flat band  $E^{(3)}_{\bm{k}}$, which originates from the destructive interference of subdimensional eigen-wavefunctions due to the lattice geometry\cite{Yin2022,Liu2020,Bergman2008}, touches the second band $E^{(2)}_{\bm{k}}$ at the center of the Brillouin zone (the $\Gamma$ point).

The band structures for $\alpha=0.8$ and $\alpha=1.2$ were obtained by solving Eq.(\ref{EnerTB}) numerically, as presented in Figs. 1(b)-(c). Similar to the situation in graphene\cite{Feilhauer2015}, when a uniaxial strain is applied in $y$ direction, the Dirac points of kagome lattice are shifted horizontally away from $K$ and $K^{\prime}$. Besides, the applied strain deforms the flat band $E^{(3)}_{\bm{k}}$, resulting in a band crossing between $E^{(2)}_{\bm{k}}$ and $E^{(3)}_{\bm{k}}$.

\begin{center}
  \vspace{+0.2cm}
  \includegraphics[width=0.95 \linewidth]{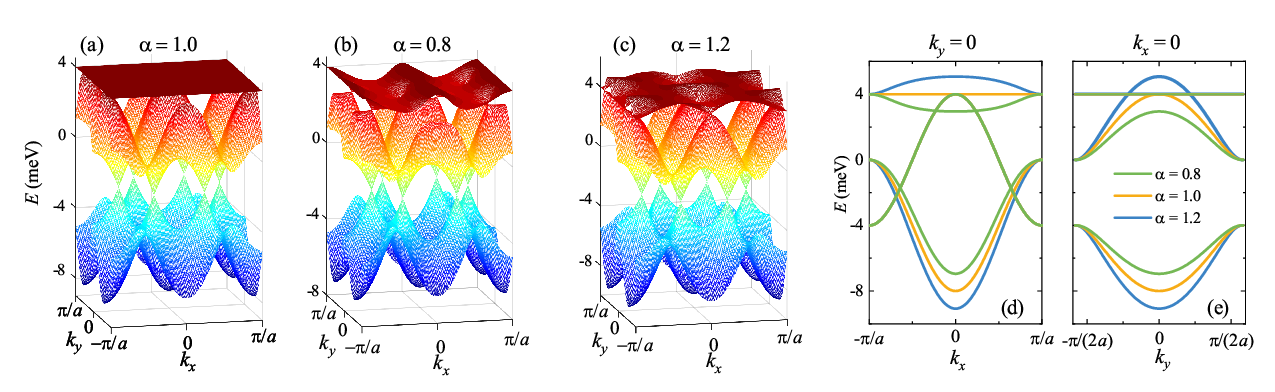}\\[10pt]
  \parbox[c]{15.0cm}{\footnotesize{\bf Fig.~2.} (Color online) (a)-(c) Band structures of strained kagome lattices with various anisotropy ratio $\alpha\equiv t^\prime/t$, calculated using a tight-binding model with only nearest-neighbour hopping. The lattice is (a) unstrained ($\alpha=1.0$), (b) stretched ($\alpha=0.8$), and (c) compressed ($\alpha=1.2$) along $y$ direction. (d)-(e) Band structures plotted along (d) the $k_x$ axis and (e) the $k_y$ axis. The hopping parameter along $x$ direction is set to $t=t_3=2$ meV.
  }\label{fig2}\\[15pt]
\end{center}

\subsubsection{Shift of Dirac points}

The shift of Dirac points are more visible in the energy bands plotted along the $k_x$ axis. As shown in Fig. 2(d), the intersection points of the two lower bands, i.e., the Dirac points, move with applied strain. The horizontal position of the two Dirac points are calculated analytically as
\begin{eqnarray}
k_x^{D_1}&=&\frac{1}{a}\arccos\left(-\frac{2\alpha}{\sqrt{8+\alpha^2}+\alpha}\right);\nonumber\\
k_x^{D_2}&=&\frac{2\pi}{a}-k_x^{D_1}.
\end{eqnarray}
Obviously, Dirac points $D_1$ and $D_2$ are symmetric with respect to the $S$ point at $k_x=\pi/a$ [see Fig. 1(b)]. Therefore, in the following we discuss only the position of $D_1$.

The movement of Dirac points in response to strain are summarized as follows. (i) For a kagome lattice stretched in $y$ direction ($\alpha<1$), the Dirac point $D_1$ moves along the $KC$ line [red lines in Fig.1(b)] towards the $C$ point at $\pi/2a$, and $D_1\rightarrow C$ when $\alpha\rightarrow 0$.
(ii) For an unstrained kagome lattice ($\alpha=1$), $D_1$ lies exactly at the $K$ point of the Brillouin zone, with $k_x^{D_1}=2\pi/(3a)$.
(iii) For a lattice subjected to a compressive strain in $y$ direction ($\alpha>1$), $D_1$ moves towards the $S$ point along the $KK^{\prime}$ line [blue lines in Fig. 1(b)], and $D_1\rightarrow S$ when $\alpha\rightarrow \infty$.

In summary, in contrast to graphene where the Dirac cones merge under applied strain\cite{Feilhauer2015}, the Dirac cones of kagome lattice remain intact at any strength of uniaxial strain within the framework of nearest-neighbor tight-binding model, indicating a more robust phase of semimetal.

\subsubsection{Reshaping of flat band}

Another notable effect of strain is the reshaping of the flat band $E^{(3)}_{\bm{k}}$. In a strained kagome lattice, the hopping anisotropy disrupts the destructive interference, leading to a deformation of the flat band, as depicted in Figs. 2(b)-(e). As a result, the intersection points of the two upper bands splits from the $\Gamma $ point and moves in opposite directions along $k_x$ direction for stretching strain ($\alpha<1$) and along $k_y$ direction for compressive strain ($\alpha>1$). The positions of the intersection points are $\alpha$-dependent, as analytically given by
\begin{eqnarray}\label{Eq7}
k_x^{I}&=&\pm\frac{1}{a}\arccos\left(\frac{2\alpha}{\sqrt{8+\alpha^2}-\alpha}\right)\, (\alpha<1);\nonumber\\
k_y^I&=&\pm\frac{1}{\sqrt{3}a}\arccos\left(\frac{2-\alpha^2}{\alpha^2}\right)\, (\alpha>1).
\end{eqnarray}
When $\alpha=0$, Eq. (\ref{Eq7}) gives $k_x^I=k_y^I=0$, indicating that the flat band $E^{(3)}_{\bm{k}}$ touches the quadratic band $E^{(2)}_{\bm{k}}$ at the $\Gamma$ point. The value of $|k_x^I|$ or $|k_y^I|$ increases with applied strain: For stretching strain, when $\alpha\rightarrow 0$, $|k_x^I|\rightarrow\pi/{2a}$, i.e., the two intersection points respectively approach $C$ and $C^{\prime}$. For compressive strain, when $\alpha\to\infty$, $|k_y^{I}|\rightarrow\pi/(\sqrt{3}a)$, i.e., the intersection points approach the $S$ point.

An intriguing feature of the deformed flat band $E^{(3)}_{\bm{k}}$ is the formation of a partially flat region that is dispersionless along $k_x$ direction whereas dispersive along $k_y$ direction, as shown in Figs. 2(b)-(e). In particular, as illustrated in Fig. 2(b), in the case of stretching strain ($\alpha<1$), this partially flat region is well separated from other energy bands and therefore becomes experimentally accessible. Due to the unconventional dispersion relation, electron states in this region possesses a highly anisotropic effective-mass tensor, in which $m_{xx}$ takes a finite value whereas $m_{yy}=\infty$. It provides a promising platform for investigating anisotropic electron transport and correlations.

\section{Artificial kagome lattice under uniaxial strain}

In this section, we numerically investigate the strain effects in artificial kagome lattices using an antidot model. Compared with the tight-binding model with only nearest-neighboring hopping terms, the antidot model based on a periodic potential is more realistic in the sense that it provides the design of antidot patterns required for experimentally fabricating artificial kagome materials from conventional two-dimensional electron gases (2DEGs)\cite{Park2009, Gibertini2009, Tadjine2016}.

\subsection{Antidot model for kagome lattice}

Two different antidot models have been proposed in the literature for realizing kagome lattices\cite{Tadjine2016, Li2016}, and here we adapt the design in Ref.\cite{Tadjine2016}. This model is based on a periodic muffin-tin potential $V(\bm{r})$ comprising two types of circular barrier regions where $V(\bm{r})$ takes a constant value $V_0>0$, as indicated by the blue circles in Fig. 3(a). The radii of the large and small barrier circles are denoted by $\rho_1$ and $\rho_2$, respectively, and the vertical spacing between nearest-neighboring circles is denoted by $L$. Most calculations in this section are performed with $\rho_1/\rho_2=3$, because such a choice ensures that the large and small barrier circles in Fig. 3(a) will touch the dashed lines simultaneously as $\rho_1$ increases. Due to the repulsion from the barriers, the wave functions of electrons mainly distribute in the space between the barrier regions, forming an artificial kagome lattice with a horizontal inter-site distance of $a$, as illustrated by the hollow circles in Fig. 3(a).

\begin{center}
  \vspace{+0.2cm}
  \includegraphics[width=0.45 \linewidth]{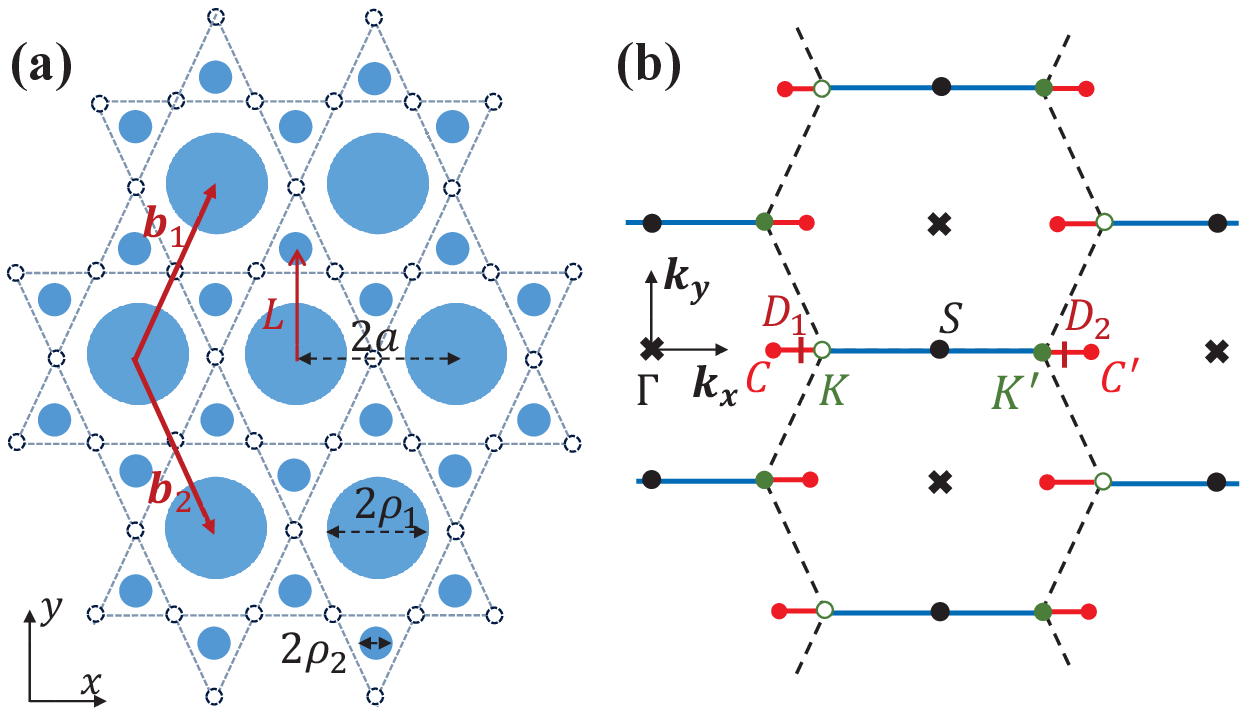}\\[5pt]
  \parbox[c]{15.0cm}{\footnotesize{\bf Fig.~3.}
  (Color online) (a) Periodic muffin-tin potential for realising artificial kagome lattice. The potential function satisfies $V(\bm{r})=V_0>0$ inside the blue circles and $V(\bm{r})=0$ elsewhere. The large solid circles form a triangular lattice spanned by the lattice vectors $\bm{b}_1$ and $\bm{b}_2$, the angle between which changes with applied strain. Each large circle is surrounded by six small circles. The hollow circles located in the gaps between blue circles illustrates the sites of kagome lattice. The lattice shown here is stretched along $y$ direction with $\alpha^{\prime}=0.8$. (b) The repeated Brillouin zone corresponds to the lattice shown in (a). The first Brillouin zone of a strained kagome lattice is a squashed hexagon. The Dirac points are located in the $CC^\prime$ line when a uniaxial strain is applied in $y$ direction.
  }\label{Fig3model}\\[15pt]
\end{center}

In the antidot model, a uniaxial strain in $y$ direction is achieved by varying the vertical spacing $L$ between adjacent barrier circles. For the unstrained case, the lattice geometry gives $L=2a/\sqrt{3}$, and the anisotropic ratio $\alpha^{\prime}$ can thus be defined as $\alpha^{\prime}\equiv2a/(\sqrt{3}L)$. Similar to the tight-binding model, here $\alpha^{\prime}=1$ represents an unstrained kagome lattice, while $\alpha^{\prime}>1$  ($\alpha^{\prime}<1$) corresponds to the situation where the lattice is compressed (stretched) in $y$ direction.

The analytical expression of $V(\bm{r})$ reads
\begin{equation}\label{potential}
V(\bm{r})=V_0 \sum_{\bm{R}}[\Theta(\rho_1-|\bm{r}-\bm{R}|)+ \Theta(\rho_2-|\bm{r}-\bm{R}\pm \bm{L}|)],
\end{equation}
where $\Theta$ is the Heaviside step function, $\bm{L}=\frac{2a}{\sqrt{3}\alpha^{\prime}}\bm{e}_y$ is the vertical displacement between adjacent barrier circles, and $\bm{R}=m \bm{b}_1+n \bm{b}_2 $ represents the lattice sites spanned by lattice vectors
\begin{equation}\label{vec2}
\bm{b}_1=a \bm{e}_x+\frac{\sqrt{3}a}{\alpha^{\prime}}\bm{e}_y  ~~\textrm{and}~~  \bm{b}_2=a \bm{e}_x-\frac{\sqrt{3}a}{\alpha^{\prime}}\bm{e}_y.
\end{equation}
Apparently, for an unstrained lattice with $\alpha^{\prime}=1$, the lattice vectors in Eq. (\ref{vec2}) reduce to those in Eq. (\ref{vec1}).

The reciprocal lattice of the lattice described by Eq. (\ref{vec2}) is given by $\bm{G}=m\bm{b}^*_1+n\bm{b}^*_2$, where
\begin{equation}
\bm{b}^*_1=\frac{\pi}{a} \bm{e}_x+\frac{\pi\alpha^{\prime}}{\sqrt{3}a}\bm{e}_y  ~~\textrm{and}~~  \bm{b}^*_2=\frac{\pi}{a} \bm{e}_x-\frac{\pi\alpha^{\prime}}{\sqrt{3}a}\bm{e}_y.
\end{equation}

In resemblance to the graphene lattice, when $\alpha^{\prime}\neq 1$, the Brillouin zone of such a lattice is an irregular hexagon squeezed along $k_x$ or $k_y$ direction, and the positions of $K$ and $K^\prime$ are
\begin{equation}
K=\left[\frac{\pi}{2a}\left( 1+\frac{\alpha^{\prime2}}{3}\right),0\right] ~~\textrm{and}~~ K^\prime=\left[\frac{\pi}{2a}\left( 3-\frac{\alpha^{\prime2}}{3}\right),0\right].
\end{equation}

\subsection{Numerical method for band calculation}
For an electron system subjected to the periodic potential $V(\bm{r})$ given by Eq.(\ref{potential}), the Bloch wave function $\psi_{\bm{k}}(\bm{r})$ satisfies the Schr\"{o}dinger equation
\begin{equation}\label{scho}
H\psi_{\bm{k}}(\bm{r})=E_{\bm{k}}\psi_{\bm{k}}(\bm{r}),
\end{equation}
where $E_{\bm{k}}$ is the eigenenergy of Bloch state $\psi_{\bm{k}}(\bm{r})$ and the Hamiltonian is $H=-\frac{\hbar^2}{2m^*}\nabla^2+V(\bm{r})$, where $m^*$ is the effective mass of electrons.

To calculate the energy bands $E_{\bm{k}}$, we expand $\psi_{\bm{k}}(\bm{r})$ with the plane-wave basis set by
\begin{equation}\label{psi}
\psi_{\bm{k}}(\bm{r})=\sum_{\bm{G}}c_{\bm{k}-\bm{G}}e^{i(\bm{k}-\bm{G})\cdot \bm{r}},
\end{equation}
where $c_{\bm{k}}$ is the coefficient of the plane wave with wave vector $\bm{k}$. Substituting Eq.(\ref{psi}) into Eq.(\ref{scho}) gives the central equation
\begin{equation}\label{central}
\left[\frac{\hbar^2\left(\bm{k}-\bm{G}\right)^2}{2m^*}-E_{\bm{k}}\right]c_{\bm{k}-\bm{G}}+\sum_{\bm{G}^\prime}V_{\bm{G}^\prime-\bm{G}}c_{\bm{k}-\bm{G}^\prime}=0,
\end{equation}
where the Fourier coefficients of $V(\bm{r})$ are given by
\begin{eqnarray}
V_{\bm{G}}&=&\frac{1}{s}\int_{cell}V(\bm{r})e^{-i\bm{G}\cdot \bm{r}}d\bm{r}\nonumber\\
&=&\frac{2\pi V_0}{s|G|}[\rho_1J_1(|G|\rho_1)+2\rho_2J_1(|G|\rho_2)\cos(\bm{G}\cdot\bm{L})].
\end{eqnarray}
Here $s=2\sqrt{3}a^2/\alpha^{\prime}$ is the area of the unit cell and $J_1$ is the Bessel of the first kind. The energy bands $E^{(i)}_{\bm{k}}$ are subsequently obtained by numerically solving Eq.(\ref{central}).

\subsection{Results and discussions}

In the following we present the numerical results obtained using the antidot model described in Sec. 3.1. The calculation was performed with $a=25$ nm and $m^*=0.04 m_e$, where $m_e$ is the free electron mass. These parameters are experimentally achievable by fabricating an antidot array onto a typical 2DEG using e-beam lithography.

\subsubsection{Width of the flat band}

The first three energy bands of the antidot model are similar to those obtained with the tight-binding model, as shown in Figs. 4(a)-(c). However, for the unstrained lattice with $\alpha^{\prime}=1$, a qualitative difference exists in the band $E^{(3)}_{\bm{k}}$ obtained using different models: the band $E^{(3)}_{\bm{k}}$, which is ideally flat within the framework of the tight-binding model, becomes weakly dispersive in the antidot model, as illustrated in Fig. 4(a). Such a dispersive band $E^{(3)}_{\bm{k}}$ in the antidot model is caused by the higher-order interactions besides the nearest-neighbour hopping.

\begin{center}
  \vspace{+0.2cm}
  \includegraphics[width=0.95 \linewidth]{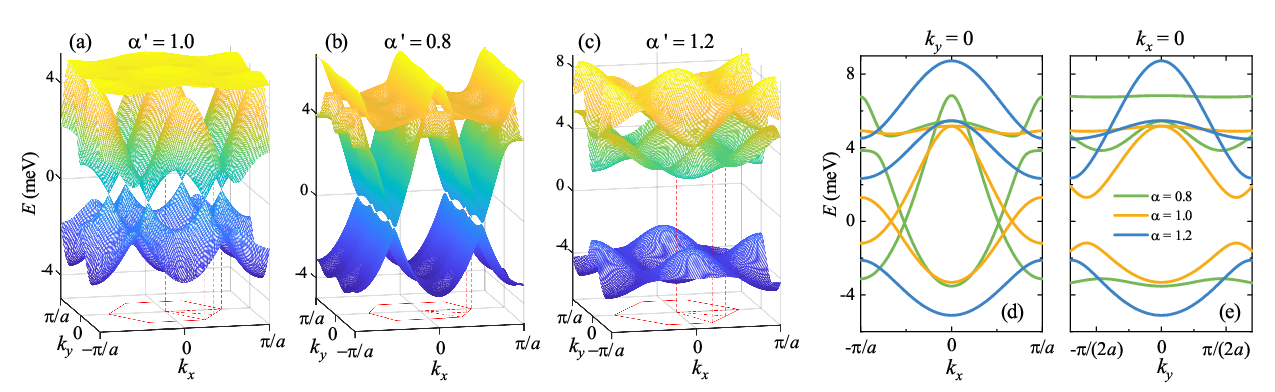}\\[5pt]
  \parbox[c]{15.0cm}{\footnotesize{\bf Fig.~4.}
  (Color online) (a)-(c) The lowest three energy bands of strained kagome lattices with various values of $\alpha^{\prime}$, calculated using the periodic muffin-tin potential described in the main text. The lattice is (a) unstrained ($\alpha^{\prime}=1.0$), (b) stretched ($\alpha^{\prime}=0.8$), and (c) compressed ($\alpha^{\prime}=1.2$) along $y$ direction. (e)-(f) Band structures plotted along (e) the $k_x$ axis and (f) the $k_y$ axis. The bands are shifted vertically to move the Dirac points to zero energy.
  All calculations were performed with $V_0 = 200$ meV, $\rho_1/\rho_2=3$ and $\beta\equiv\rho_1/a=0.6$.
  }\label{Fig5}\\[15pt]
\end{center}

The band width $\Delta E$ of $E^{(3)}_{\bm{k}}$ can be tuned by adjusting the parameters of the antidot model. To test how $\Delta E$ depends on the model parameters, we calculated $\Delta E$ as a function of barrier height $V_0$ with various values of $\beta\equiv\rho_1/a$, as presented in Fig. 5.

\begin{center}
  \vspace{+0.2cm}
  \includegraphics[width=0.4 \linewidth]{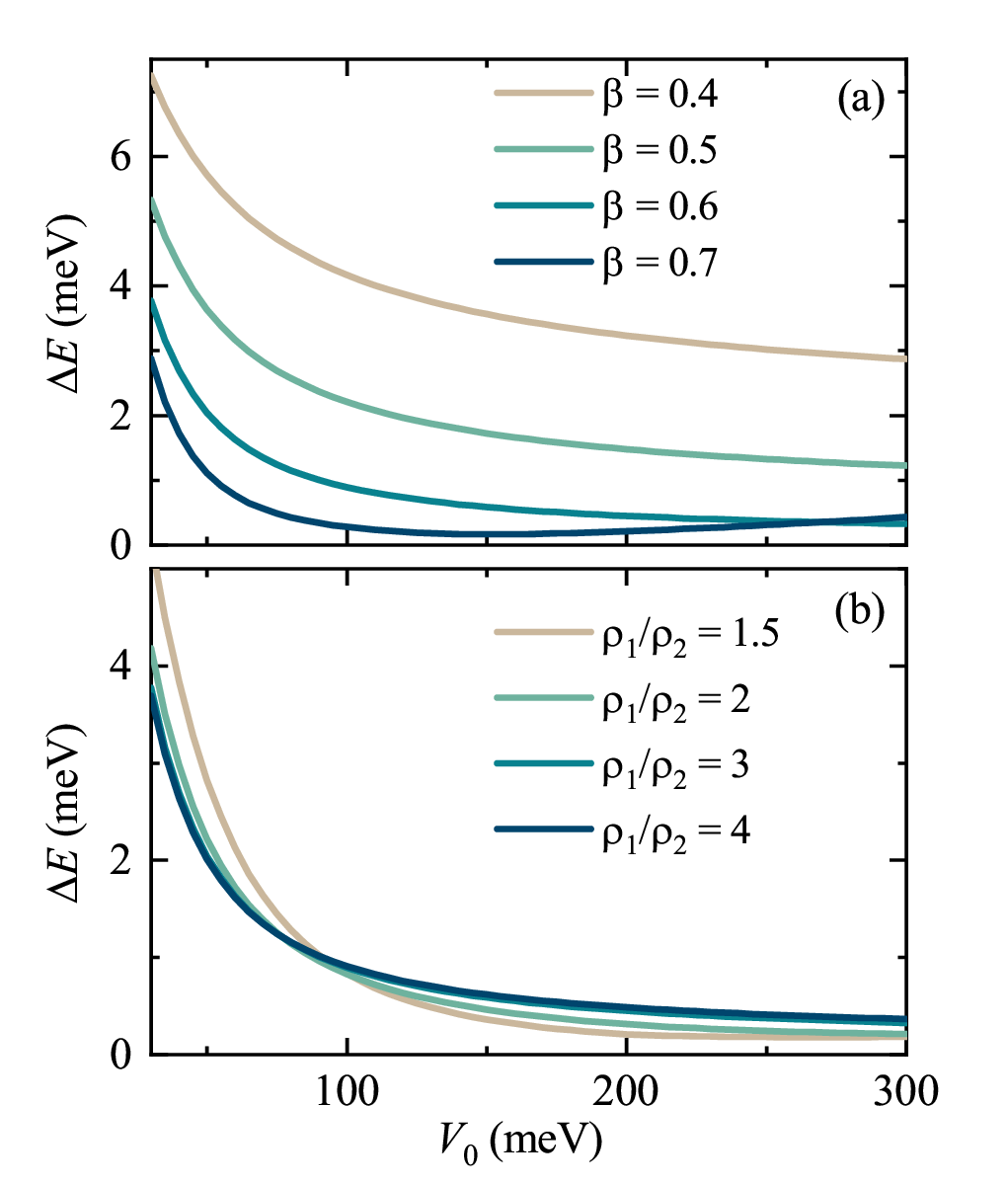}\\[5pt]
  \parbox[c]{15.0cm}{\footnotesize{\bf Fig.~5.} (Color online) Band width $\Delta E$ of the flat band $E^{(3)}_{\bm{k}}$ versus barrier height $V_0$ of the maffin-tin potential $V(\bm{r})$ calculated with (a) $\rho_1/\rho_2=3$ and various values of $\beta\equiv\rho_1/a$, and (b) various values of $\rho_1/\rho_2$ and $\beta=0.6$.
  }\label{Fig4}\\[15pt]
\end{center}

When either $V_0$ or $\beta$ is small, the wave functions of electrons are not well localized to the regions of lattice sites illustrated by the hollow circles in Fig. 3(a). In this situation, the antidot model cannot be well approximated by a tight-binding lattice model and therefore the obtained $\Delta E$ of $E^{(3)}_{\bm{k}}$ is large.
As shown in Fig. 5, $\Delta E$ generally decreases with increasing $V_0$ and $\beta$ except for the high $V_0$ region of the curve with $\beta=0.7$. A small band width of $\Delta E=0.45$ meV is obtained with parameters $\beta=0.6$ and $V_0=200$ meV; the calculations presented in Fig. 4 were all carried out with this group of parameters.

The band width $\Delta E$ of $E^{(3)}_{\bm{k}}$ is also slightly influenced by the ratio $\rho_1/\rho_2$. As shown in Fig. 5(b), $\Delta E$ increases with increasing $\rho_1/\rho_2$ for small values of $V_0$, whereas $\Delta E$ decreases with increasing $\rho_1/\rho_2$ for large values of $V_0$.

\subsubsection{Merging of Dirac points}

As discussed in Sec. 2, when a kagome lattice is strained in $y$ direction, the tight-binding model predicts that the Dirac cones will move horizontally but will never merge. However, as illustrated in Figs. 4(c)-(e), the numerical results of antidot model clearly demonstrate the merging of Dirac cones under strong compressive strain, resulting in a band gap opening between bands $E^{(1)}_{\bm{k}}$ and $E^{(2)}_{\bm{k}}$.

To quantitatively present the movement of the Dirac points in response to applied strain, we introduce a parameter $\gamma$ to describe the relative position of the Dirac points $D_1$ and $D_2$ with respect to the $S$ point [see Fig. 3(b)]. The parameter $\gamma$ is defined as
\begin{equation}\label{gamma}
\gamma\equiv\frac{|\bm{k}^{D_1}-\bm{k}^{S}|}{|\bm{k}^{C}-\bm{k}^{S}|}=\frac{|\bm{k}^{D_2}-\bm{k}^{S}|}{|\bm{k}^{C^{\prime}}-\bm{k}^{S}|},
\end{equation}
where $|\bm{k}^{C}-\bm{k}^{S}|=|\bm{k}^{C^{\prime}}-\bm{k}^{S}|=\pi/a$. Obviousely, $\gamma=1$ corresponds to the situation in which the Dirac points $D_1$ and $D_2$ are respectively shifted to $C$ and $C^\prime$, and $\gamma=0$ is reached when $D_1$ and $D_2$ meet and merge at the $S$ point. In addition, in the antidot model, it is noteworthy that both $K$ and $K^{\prime}$ move with applied strain due to the deformation of the lattice. Therefore, a parameter $\gamma_K$ is similarly introduced to describe the relative position of $K$, defined as
\begin{equation}\label{gammaK}
\gamma_K=\frac{|\bm{k}^{K_1}-\bm{k}^{S}|}{|\bm{k}^{C}-\bm{k}^{S}|}=1-\frac{\alpha^{\prime2}}{3}.
\end{equation}
When $\alpha^{\prime}=\sqrt{3}$, Eq. (\ref{gammaK}) yields $\gamma_K=0$, indicating the merging of $K$, $K^\prime$ and $S$.

\begin{center}
  \vspace{+0.2cm}
  \includegraphics[width=0.45 \linewidth]{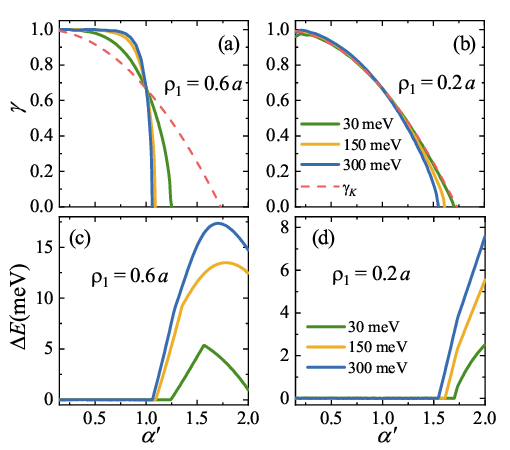}\\[5pt]
  \parbox[c]{15.0cm}{\footnotesize{\bf Fig.~6.} (Color online) (a, b) Relative position $\gamma$ of Dirac points as a function of anisotropy ratio $\alpha^{\prime}$ for (a) $\beta\equiv\rho_1/a=0.6$ and (b) $\beta=0.2$, calculated with $\rho_1/\rho_2=3$ and various values of $V_0$. The dashed lines depict the position of the $K$ point given by Eq.(\ref{gammaK}). (c, d) Band gap $\Delta E$ at Dirac points as a function of $\alpha^{\prime}$, calculated with (c) $\beta=0.6$ and (d) $\beta=0.2$.
  }\label{Fig6}\\[15pt]
\end{center}

The relative position $\gamma$ of the Dirac points were plotted against $\alpha^{\prime}$ with various values of $V_0$ and $\beta$, as presented in Figs. 6(a)-(b). When the artificial lattice is subjected to a stretching strain along $y$ direction ($\alpha^{\prime}<1$), the Dirac points $D_1$ and $D_2$ moves respectively towards $C$ and $C^{\prime}$ ($\gamma\to1)$. On the other hand, with an increasing compressive strain ($\alpha^{\prime}>1$), both $D_1$ and $D_2$ move towards $S$ ($\gamma\to0$) and finally merge at $S$ ($\gamma=0$). As shown in Figs. 6(c)-(d), right after the merging of the Dirac cones, a band gap opens between $E^{(1)}_{\bm{k}}$ and $E^{(2)}_{\bm{k}}$, indicating the occurrence of a phase transition from a semimetal to a semiconductor.

\subsubsection{Partially flat regions in deformed flat bands}

The antidot model also predicts a deformation of the flat band $E^{(3)}_{\bm{k}}$ when the lattice is strained in $y$ direction, as shown in Figs. 4(b)-(e). In the unstrained case ($\alpha^{\prime}=1$), band $E^{(2)}_{\bm{k}}$ touches $E^{(3)}_{\bm{k}}$ at the $\Gamma$ point. With an increasing stretching strain applied in $y$ direction ($\alpha^{\prime}<1$), bands $E^{(2)}_{\bm{k}}$ and $E^{(3)}_{\bm{k}}$ cross at two points on the $k_x$ axis, and the intersection points move in opposite directions toward $C$ and $C^\prime$, respectively. On the contrary, a compression in $y$ direction makes the two crossing points on the $k_y$ axis shift respectively along $\pm k_y$ direction toward the edges of the first Brillouin zone. Such a strain-induced deformation of the flat band is in consistency with the results of the tight-binding model.

\begin{center}
  \vspace{+0.2cm}
  \includegraphics[width=0.35 \linewidth]{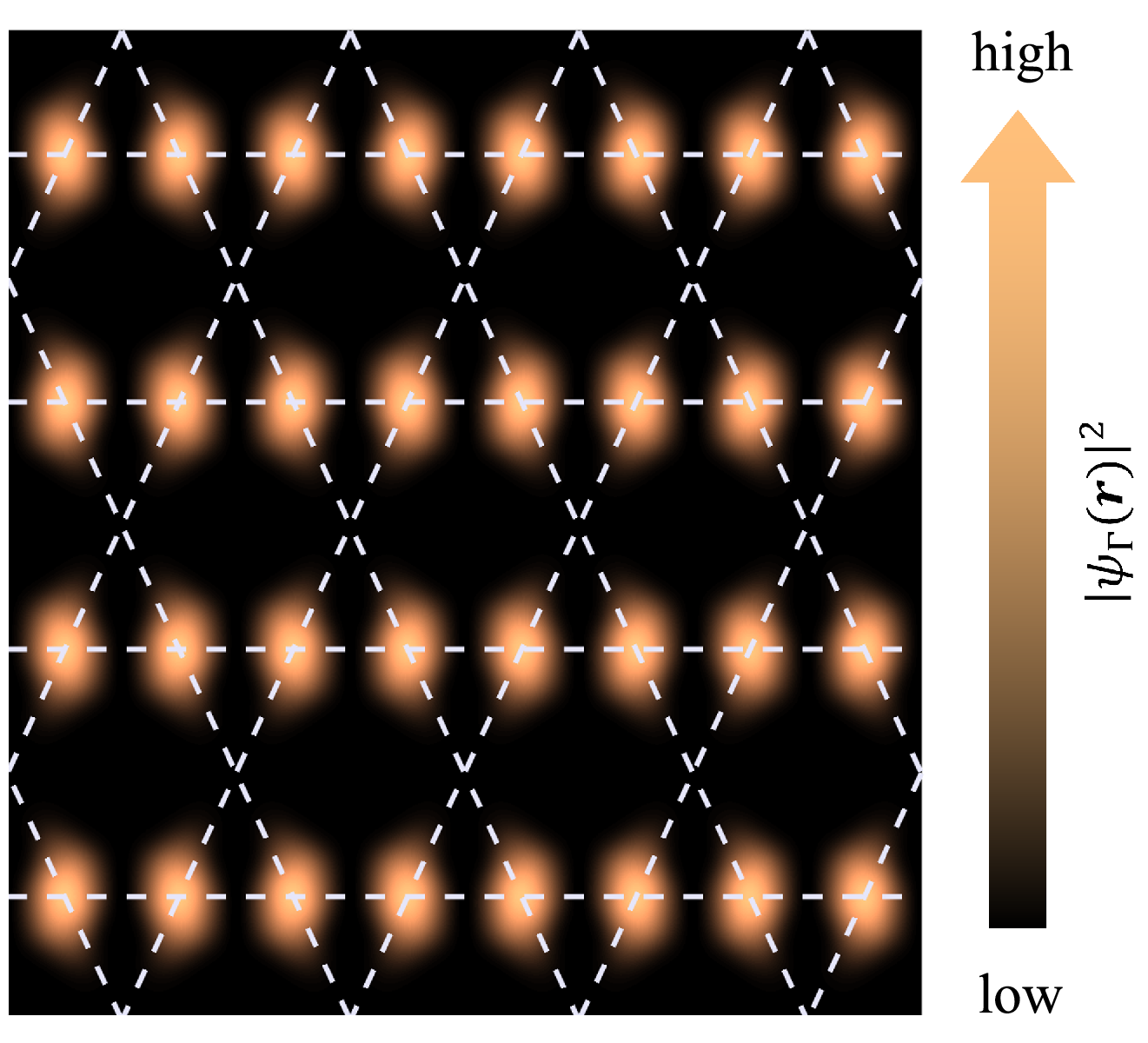} \\[5pt]
  \parbox[c]{15.0cm}{\footnotesize{\bf Fig.~7.} (Color online) Probability density distribution of the Bloch state $\psi_{\Gamma}(\bm{r})$ of band $E^{(3)}_{\bm{k}}$, calculated with $\alpha^{\prime}=0.8$. Dashed lines illustrate the kagome lattice. The parameters of muffin-tin potential are the same to those in Fig.5.
  }\label{Fig7}\\[15pt]
\end{center}

Similar to the results of the tight-binding model, a partially flat region well separated from other bands is also found in the deformed flat band $E^{(3)}_{\bm{k}}$ with $\alpha^{\prime}=0.8$. This highly anisotropic region is almost dispersionless along $k_x$ direction whereas clearly dispersive along $k_y$ direction, as shown in Figs. 4(b) and (e). To gain further insight into the origin of the band anisotropy of this region, we calculated the real-space probability density distribution of the Bloch state $\psi_{\Gamma}(\bm{r})$ of band $E^{(3)}_{\bm{k}}$, as presented in Fig. 7. It is found that the wave function $\psi_{\Gamma}(\bm{r})$ mainly distribute around sites $B$ and $C$ in the unit cell while leave the site $A$ empty. Such an anisotropic site occupation is consistent with the anisotropic effective mass tensor obtained from band calculation.

\section{Further discussions}

\subsection{Band structures along high-symmetry paths}

To better illustrate the evolution of the band features in response to applied strain, we plot the band-structure data calculated using both the tight-binding model and the antidot model along the high-symmetry path $\Gamma$-$M$-$K$-$\Gamma$-$M^{\prime}$-$K^{\prime}$. As shown in Fig. 8, the band structures calculated using the tight-binding model always show characteristic van Hove singularities at both the $M$ and $M^{\prime}$ points, irrespective of applied strain. However, for the band structures calculated using the antidot model, the van Hove singularity at the $M$ point disappears when $\alpha^{\prime}=0.8$ whereas that at the $M^{\prime}$ point disappears when $\alpha^{\prime}=1.2$, as illustrated in Figs. 9(b)-(c).

\begin{center}
  \vspace{+0.2cm}
  \includegraphics[width=0.4 \linewidth]{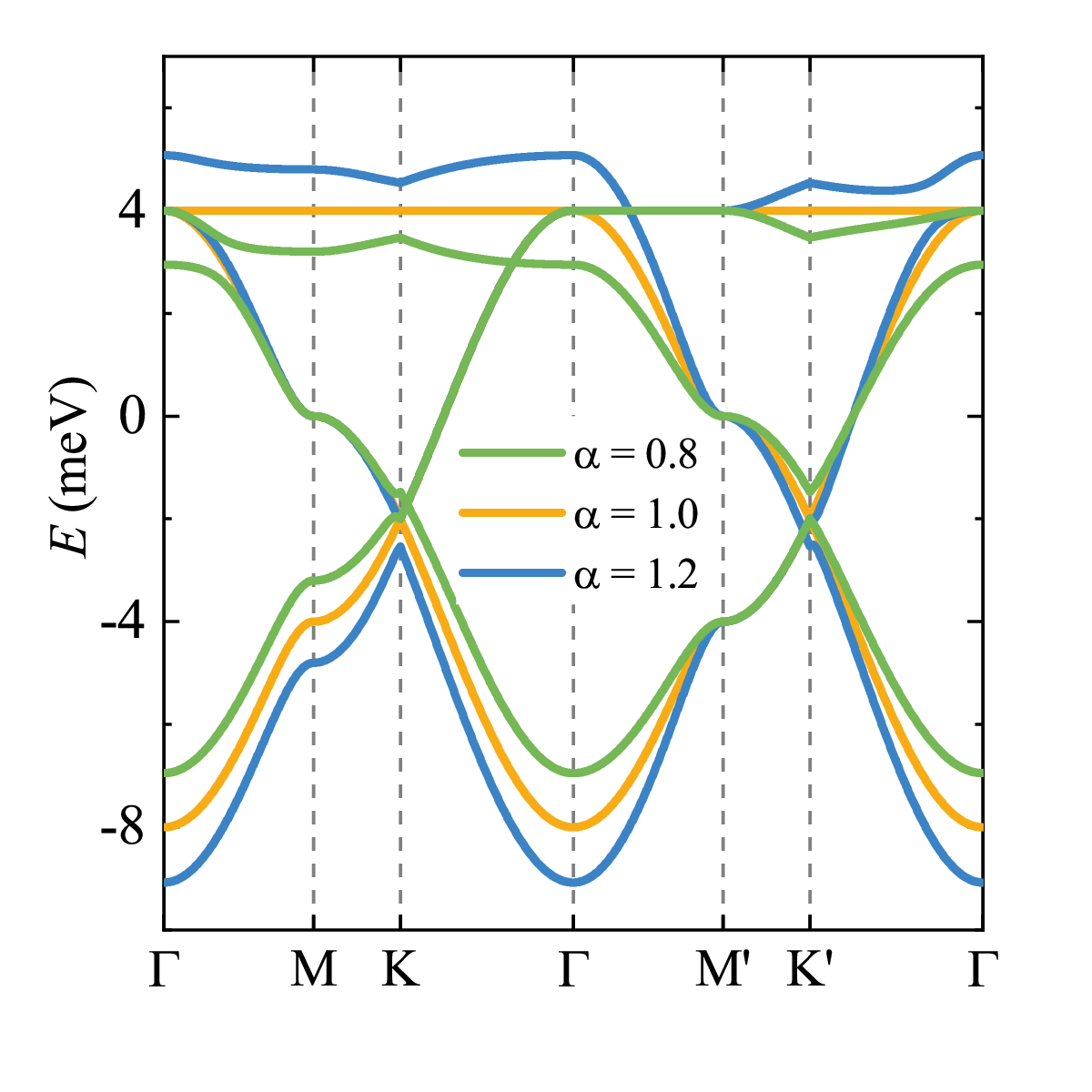} \\[5pt]
  \parbox[c]{15.0cm}{\footnotesize{\bf Fig.~8.} (Color online) Plots of the band-structure data in Fig. 2 along the high-symmetry path. The corresponding Brillouin zone is illustrated in the left panel of Fig. 9(a).
  }\label{Fig8}\\[15pt]
\end{center}

\begin{center}
  \vspace{+0.2cm}
  \includegraphics[width=0.4 \linewidth]{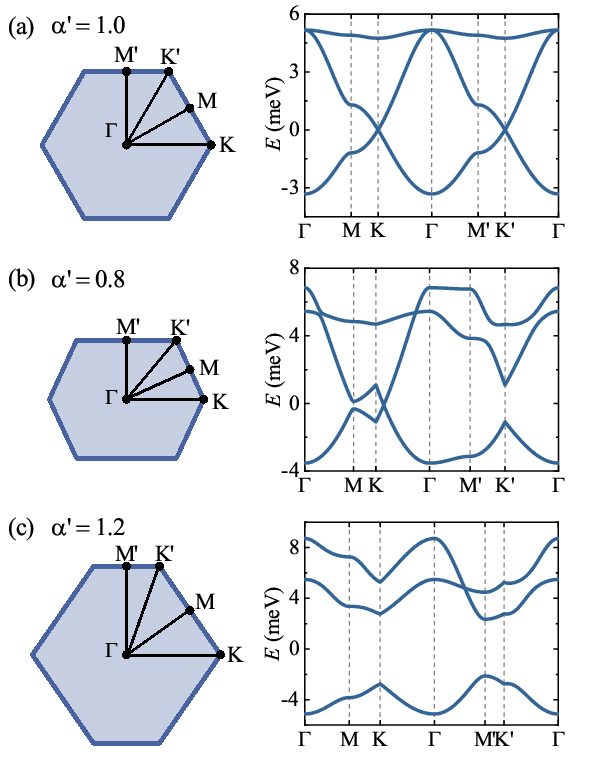} \\[5pt]
  \parbox[c]{15.0cm}{\footnotesize{\bf Fig.~9.} (Color online) Plots of the band-structure data in Fig. 4 along the high-symmetry path when the lattice is (a) unstrained ($\alpha^{\prime}=1.0$), (b) stretched ($\alpha^{\prime}=0.8$), and (c) compressed ($\alpha^{\prime}=1.2$) along $y$ direction. The corresponding Brillouin zones are shown on the left-hand side of the data.
  }\label{Fig9}\\[15pt]
\end{center}

\subsection{Comparison between the two models}

As presented in the previous sections, discernible differences were found between the calculation results based on the nearest-neighboring tight-binding model and the antidot model. The main differences include: (i) The band $E^{(3)}_{\bm{k}}$ of the unstrained kagome lattice is ideally flat in the tight-binding model whereas it is slightly dispersive in the antidot model. (ii) The merging of Dirac cones and the opening of band gaps between $E^{(1)}_{\bm{k}}$ and $E^{(2)}_{\bm{k}}$ due to compressive strains occurs only in the antidot model. (iii) When the lattice is subjected to applied strain, the two van Hove singularities at $M$ and $M^{\prime}$ remain unchanged in the tight-binding model whereas one of them disappears in the antidot model. 

In the following, we compare the two theoretical models and briefly discuss their advantages and limitations. The nearest-neighboring tight-binding model is used for approximating a periodic potential $V(\bm{r})$  in which electrons are strongly bound to the lattice sites. It assumes that the energy eigenstate $\phi_i$ of the $i^{\textrm{th}}$ isolated site distributes only in a small region near the site and therefore the hopping parameters $t_{ij}=\langle\phi_{i}|\hat{H}|\phi_j\rangle$ take non-zero values only between neighboring sites $i$ and $j$. The advantage of the nearest-neighboring tight-binding model lies in its mathematical simplicity: in many situations, the energy band $E^{(i)}_{\bm{k}}$ can be obtained analytically as a function of $t_{ij}$. In this work, the analytical results obtained using the tight-binding method helps us to intuitively understand the evolution of band structures under strain. On the other hand, the obvious limitation of this model is that it cannot nicely describe periodic potentials in which electrons are not firmly bound to the lattice sites and thus the hopping between non-adjacent lattice sites cannot be neglected.

The antidot model based on a periodic muffin-tin potential is specifically designed for modeling the artificial superlattice realized by patterning an array of holes (called an antidot array) on a 2DEG. In the antidot model, the potential function $V(\bm{r})$ is explicitly given. In this regard, it is more realistic than the tight-binding model. However, the application of antidot model is limited to artificial kagome superlattices and it is not directly applicable to kagome materials and other artificial kagome systems.

As discussed above, these two theoretical models are not mathematically equivalent. The antidot model given by Eq.(8) cannot be perfectly mapped to a nearest-neighboring tight-binding model because electrons in the antidot model are not bound firmly enough to the lattice sites shown in Fig. 3(a). Such a mathematical inequivalence is the origin of the differences in the calculation results obtained using the two models.

\section{Conclusion}

In summary, we have performed a comprehensive study on the band structures of strained kagome lattices using both a tight-binding model and an antidot model. Both models predict a strain-induced horizontal shift of the Dirac cones in the band structures. In addition, according to the antidot model, when the lattice is subjected to a strong compressive strain in $y$ direction, the Dirac cones will merge and a band gap will open between the two lowest energy bands. Furthermore, in a kagome lattice stretched along $y$ direction, the flat band $E^{(3)}_{\bm{k}}$ is found to develop into a highly anisotropic shape, with a partially flat region dispersionless along $k_y$ direction while dispersive along $k_x$ direction. Our results pave the way for engineering the electronic band structures of kagome materials by mechanical strain.

\vspace{1.5cm}
{\flushleft\it Acknowledgements.}---
This work was supported by the National Natural Science Foundation of China (Grants Nos.\ 11904261 \& 11904259).


\end{document}